\PassOptionsToPackage{table, dvipsnames}{xcolor}

\documentclass[sigconf, authorversion, noacm]{acmart}

\usepackage{mystyle}

\def\doctitle{Synthetic End-User Testing:\\Modeling Realistic Agents Based on Behavioral Examples}

\def\docauthors{Pasquale Salza, Marco Edoardo Palma, Harald C. Gall}

\def\dockeywords{%
Multi-agent systems, agents, testing, neural networks, Markov models, deep learning
}

\StrSubstitute{\doctitle}{\\}{ }[\cleandoctitle]
\StrSubstitute{\dockeywords}{.}{}[\cleandockeywords]
\hypersetup{
	pdftitle={\cleandoctitle},
	pdfauthor={\docauthors},
	pdfkeywords={\cleandockeywords}
}

\def\replicationurl{https://github.com/MEPalma/SyntheticEndUserTesting-Prototype}

\DeclareAcronym{api}{
	short = API,
	long = {Application Program Interface}
}

\DeclareAcronym{mas}{
	short = MAS,
	long = {Multi-Agent Systems}
}

\DeclareAcronym{mcmc}{
	short = MCMC,
	long = {Markov Chain Monte Carl}
}

\DeclareAcronym{mlm}{
	short = MLM,
	long = {Masked Language Model}
}

\DeclareAcronym{nsp}{
	short = NSP,
	long = {Next Sentence Prediction}
}

\DeclareAcronym{bert}{
	short = BERT,
	long = {Bidirectional Encoder Representations from Transformers}
}

\DeclareAcronym{ast}{
	short = AST,
	long = {Abstract Syntax Tree}
}

\DeclareAcronym{cfg}{
	short = CFG,
	long = {Control Flow Graph}
}

\DeclareAcronym{gpt}{
	short = GPT,
	long = {Generative Pretrained Transformer}
}

\DeclareAcronym{ir}{
	short = IR,
	long = {Information Retrieval}
}

\DeclareAcronym{lstm}{
	short = LSTM,
	long = {Long Short-Term Memory}
}

\DeclareAcronym{nn}{
	short = nn,
	long = {Neural Network}
}

\DeclareAcronym{rnn}{
	short = RNN,
	long = {Recurrent Neural Network}
}

\DeclareAcronym{brnn}{
	short = BRNN,
	long = {Bidirectional Recurrent Neural Network}
}

\DeclareAcronym{cnn}{
	short = CNN,
	long = {Convolutional Neural Network}
}

\DeclareAcronym{tf-idf}{
	short = tf-idf,
	long = {term frequency–-inverse document frequency}
}

\DeclareAcronym{anova}{
	short = ANOVA,
	long = {ANalysis Of VAriance}
}

\DeclareAcronym{ta}{
	short = TA,
	long = {Task-Adaptive}
}

\DeclareAcronym{gru}{
	short = GRU,
	long = {Gated Recurrent Unit}
}

\DeclareAcronym{gan}{
	short = GAN,
	long = {Generative Adversarial Networks}
}

\DeclareAcronym{sh}{
	short = SH,
	long = {Syntax Highlighting}
}

\DeclareAcronym{re}{
	short = Regex,
	long = {Regular Expressions}
}

\DeclareAcronym{om}{
	short = OM,
	long = {Oracle Methods}
}

\DeclareAcronym{bf}{
	short = BF,
	long = {Brute-Force}
}

\DeclareAcronym{eta}{
	short = ETA,
	long = {Extended Token Annotation}
}

\DeclareAcronym{heta}{
	short = HETA,
	long = {Highlighted Extended Token Annotation}
}

\DeclareAcronym{ide}{
	short = IDE,
	long = {Integrated Development Environment}
}

\DeclareAcronym{ui}{
	short = UI,
	long = {User Interface}
}

\DeclareAcronym{gui}{
	short = GUI,
	long = {Graphical User Interface}
}

\DeclareAcronym{aop}{
	short = AOP,
	long = {Aspect-Oriented Programming}
}

\DeclareAcronym{mvc}{
	short = MVC,
	long = {Model-View-Controller}
}

\begin{document}

\title[\cleandoctitle]{\doctitle}

\author{Pasquale Salza}
\orcid{0000-0002-8687-052X}
\affiliation{%
	\institution{University of Zurich}
	\country{Switzerland}
}
\email{salza@ifi.uzh.ch}

\author{Marco Edoardo Palma}
\orcid{0000-0003-3300-4828}
\affiliation{%
	\institution{University of Zurich}
	\country{Switzerland}
}
\email{marcoepalma@ifi.uzh.ch}

\author{Harald C. Gall}
\orcid{0000-0002-3874-5628}
\affiliation{%
	\institution{University of Zurich}
	\country{Switzerland}
}
\email{gall@ifi.uzh.ch}

\renewcommand{\shortauthors}{Salza et al.}

\begin{abstract}
For software interacting directly with real-world end-users, it is common practice to script scenario tests validating the system's compliance with a number of its features.
However, these do not accommodate the replication of the type of end-user activity to which the system is required to respond in a live instance.
It is especially true as compliance might also break in scenarios of interactions with external events or processes, such as other users.
State-of-the-art approaches aim at inducing the software into runtime errors by generating tests that maximize some target metrics, such as code coverage.
As a result, they suffer from targeting an infinitely large search space, are severely limited in recognizing error states that do not result in runtime errors, and the test cases they generate are often challenging to interpret.
Other forms of testing, such as \emph{Record-Replay}, instead fail to capture the end-users' decision-making process, hence producing largely scripted test scenarios.
Therefore, it is impossible to test a software's compliance with unknown but otherwise plausible states.
This paper introduces \enquote{Synthetic End-User Testing,} a novel testing strategy for complex systems in which real-world users are synthesized into reusable agents and employed to test and validate the software in a simulation environment.
Hence, it discusses how end-user behavioral examples can be obtained and used to create agents that operate the target software in a reduced search space of likely action sequences.
The notion of action expectation, which allows agents to assert the learned compliance of the system, is also introduced.
Finally, a prototype asserting the feasibility of such a strategy is presented.
\end{abstract}



\setcopyright{none}
\settopmatter{printacmref=false}
\renewcommand\footnotetextcopyrightpermission[1]{}
\settopmatter{printfolios=true}
\pagestyle{plain}

\maketitle

\section{Introduction}
\label{sec:introduction}

Nowadays, highly popular and complex systems that enable the interaction between multiple users, \eg, social platforms, pose a unique and non-trivial challenge in software testing~\cite{ahlgren_wes_2020}.
Such systems might be affected by issues that only arise under real-world interactions between multiple users, such as problems of security, integrity, and privacy~\cite{ahlgren_wes_2020}.
For this reason, today's testing methods might appear limited.
Traditional testing strategies overpower the developer's role, who is expected to know the popular use patterns of the system's users and needs to compose tests for all such scenarios.
Instead, state-of-the-art testing approaches, \eg, \emph{Search-Based Software Testing}, generally optimize on code metrics, \eg, coverage, which lack correlation with actual end-user behavior~\cite{mcminn_searchbased_2004}.

One way of testing would be to use real end-users.
Indeed, the system is likely to be already running in another production instance, with a community of real users interacting with it and each other.
Alternatively, the platform may still be at a development status for which it would at least be possible to test it with a restricted group of testers.
On the one hand, it is risky to mix testing activities with production systems.
On the other hand, using human testers in an isolated development environment can hardly resemble a realistic system load.
Therefore, it is paramount that the systems are tested with a simulated user activity that resembles sequences of user requests found in production.
This raises two main concerns:
\begin{inparaenum}
    \item how can a simulation environment be constructed to execute such tests?
    \item how can realistic user activity be created and tested?
\end{inparaenum}

Existing approaches have been focusing on recreating a model of the target platform, on which a simulation can be run~\cite{ie_recsim_2019}.
However, generally, platforms are complex to the point that manually compiling accurate models of user behavior becomes a tedious task and intractable, with results far from the expected use patterns experienced in the target live instance.
Therefore, such simulations and models remain at a high level of abstraction and are often far from reflecting a realistic scenario of a production system.
One may argue that the manual scripting of test scenarios might be avoided if current state-of-the-art approaches in the field of software test generation were adapted to operate as users.
Baseline solutions in this space, such as \emph{Monkey Testing}~\cite{monkey} are limited to generating user activity by triggering random actions on the client and not effective in reproducing user activity~\cite{paydar_empirical_2020, mohammed_empirical_2019}.
However, while this would potentially test for all possible activity sequences the system may undergo in production, it also tries within a much greater activity set, most of which are highly unlikely to occur.
Moreover, the results of multiple random testers operating in parallel on the system might be unpredictable and uncontrollable.
Instead, advanced solutions inform the action selection process through some internally evolving logic that strives to operate the target software in ever more effective ways to discover control sequences leading to some runtime error~\cite{perera_experimental_2022, fraser_whole_2013}.

Some of these strategies also explicitly target \ac{gui}~\cite{mao_sapienz_2016} software.
However, these are in a constant evolution of an outer process focused on boosting the strategy's effectiveness in maximizing some software testing-related metrics, \eg, the coverage, or a combination of such metrics.
As a result, these approaches typically lead to the generation of test scenarios that are difficult to comprehend and part of a more extensive search space that includes tests largely unaware of the behavior or intended/expected use patterns of the software they test.
Moreover, tests generated through such strategies are known to be largely ineffective in the detection of bugs~\cite{almasi_industrial_2017, shamshiri_automatically_2015}.

In light of the challenges raised in this space, this work introduces \emph{Synthetic End-User Testing}, a novel strategy for the automated testing of complex systems, in which real-world end-user are simulated by agents~\cite{wooldridge_intelligent_1999} that operate on an instance of the target system.
The test scenario search space can be reduced to one consisting of the set of use patterns that users of the system in production have shown to perform.
It is a desirable proposition, as it will validate a system's compliance to the set of most likely use cases, redirecting development time towards changes improving real-world use cases.
Intuitively, the information for exploring this search space can be inferred by the systems' current user base.
For the average case, there exists the opportunity to track both the actions and interactions of real users and how the state of the target platform changes over time.
Therefore, in this context, generating test scenarios moves the focus away from generators that maximize the test coverage but instead towards minimizing the distance of synthetic end-user models to their respective and unknown end-user processes.
Furthermore, current system compliance represents valuable information for encoding user expectations following some action.
Modeling user expectations of the system's behavior would also allow for the statistical assertion of the system operation.

The proposed approach consists of three components, \ie, track, synthesize, and play.
They are in charge of registering end-user activities, then encode into automated agents, and finally reproducing actions into a simulated environment.
Each component opens to many strategies, technologies, and methods that represent the primary goal envisioned in this paper.
The tracking might be implemented for different programming languages and frameworks.
The synthesis of agents might be based on statistical methods or using neural networks trained on the mole of collected user activities.
Finally, the reproduction of agents can target different testing goals, such as stress testing and bug discovery.

The remainder of the paper provides an overview of the approach and its components and presents a prototype of this novel approach to support its feasibility.
The replication package is published at the address \url{\replicationurl}~\cite{replicationpackage}.

\section{Approach}
\label{sec:approach}

\begin{figure}[tb]
    \centering
    \includegraphics[width=1.0\linewidth]{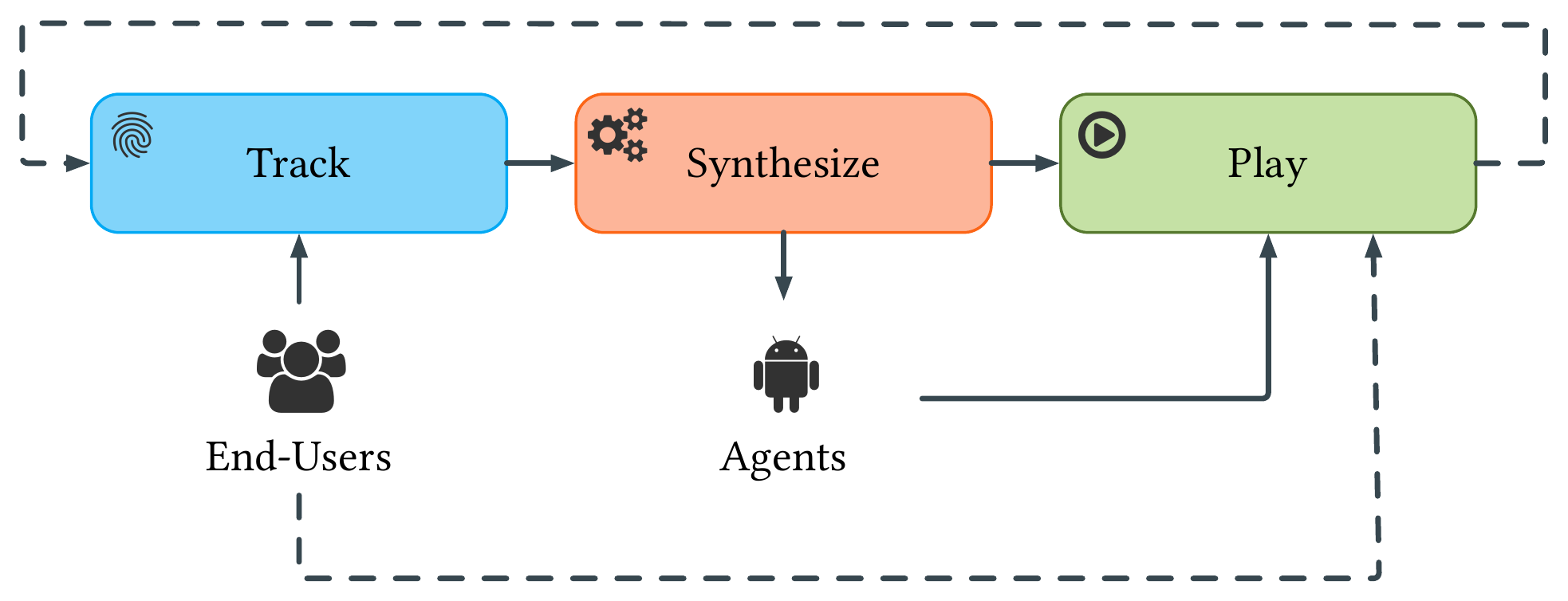}
    \caption{The workflow of the approach.}
    \label{fig:approach:workflow}
\end{figure}

The proposed approach consists of three different components, \ie, track, synthesize, and play.
\Cref{fig:approach:workflow} shows the workflow of how such elements are connected and interact with the \enquote{real} and mock actors.
In the vision of \emph{Synthetic End-User Testing}, each component is highly cohesive but weakly coupled, and different technologies and methods can be combined to reach various goals for testing.
For instance, one might decide to
\begin{inparaenum}
    \item \emph{track} end-users of an \textsc{Android} client for a social network,
    \item use statistical methods to \emph{synthesize} automated agents, then
    \item \emph{play} the agents to stress the platform under test to discover crashes or check whether the performance decays.
\end{inparaenum}
In the following, each of the components is discussed.

\paragraph{Track}
The first phase consists of having real end-users \emph{using} the target application while their actions are being tracked.
The user activity is registered as a sequence of bindings of application states and user actions.
The data recorded must be sufficient for modeling user behavior regarding the application's state.
An agent should recognize the state the application is in and, by inference from the behavior the user had previously shown to take, sample one action out of those made available to the user.
The representation of the application state should also enable the agent to validate the consequences of an action it has taken.
It means that if the application had consistently promoted responses of some specific trend to the same subsequence of actions, the agent should be able to assert if the application is still compliant with that behavior.
A tracking layer should also enable the agent to perform an action in a manner that promotes the application to carry out the exact computation as if a real user had taken it.
Tracking has to happen transparently to the end-user without showing any decay of the user experience.

Developers have to decide what and when to track, therefore, they instrument the original program code that ideally has to remain intact.
This part is envisioned to be technology-dependent since a dedicated tracker has to be specifically developed for the technology employed for the clients.
If a specific programming language or framework is used, there might be the possibility of leveraging some convenient functionalities.
For instance, the user interface of an \textsc{Android} app is likely based on a widget toolkit library that manages the \ac{gui}.
Furthermore, a tracking procedure might then be injected into the library without the need to touch the original codebase through technologies, \eg, \ac{aop}.
It is up to the developers to decide the level of granularity of such tracking.
This choice will then be relevant during the next phase of synthesis.

\paragraph{Synthesize}
Once track(s) of individual end-users have been recorded, it is time to model the agents.
An agent is meant to be a synthetic encoding of the user behavior its target end-user has led to observe.
As black boxes, agent models are given user tracking data, then acting as predictors of the following action given the application state, set of available actions, and some internal state.
The leveraging of end-user activity opens the door to a number of candidate strategies for the encoding of user processes.

A baseline approach might assume that the variability of action selection for a given user is only dependent on the system's current state.
Hence, user data might be compiled into baseline models such as Markov Chain~\cite{gagniuc_markov_2017, ibe_special_2014}, in which the system states are nodes of the decision graph.
A set of actions are sampled with the observed frequency probability, a procedure that, in turn, semantically leads the user to request the system to transition from one state to another.

Alternatively, inspiration from recent leaps in the building of language models~\cite{vaswani_attention_2017, devlin_bert_2019} may be considered for the generation of user action models.
The training of language models in \ac{mlm} and \ac{nsp}~\cite{vaswani_attention_2017} share many similarities with user action modeling.
An initial approach may consider \emph{masked words} to be \emph{masked actions}, allowing the model to learn the bidirectional structure of action sequences, while the recognition of ordered entailment between sets of action sequences might promote the learning of the semantics, or simulate the context, for which two or more such sequences follow each other.

Finally, the opportunity to produce generative user action models should be considered with approaches such as \ac{gan}~\cite{goodfellow_generative_2014} models.
Similarly, information about the system's response, or resulting state, to user actions might be used for the modeling of user expectations.
For this purpose, deep auto-encoder models used in anomaly detection may be adapted to this problem~\cite{an_variational_2015, zhou_anomaly_2017, beggel_robust_2019}.
If trained to reconstruct the system's state following a user's action, future reconstructions may signal the level of surprise about the system's operation.

\paragraph{Play}
The generated agents are now ready to be deployed into a testing environment.
\emph{Synthetic End-User Testing} is not necessarily tied to a specific testing goal.
One might decide to test the system's performance under stress through the mean of several multiple realistic users or discover bugs that derive from the simultaneous use of the system, which are otherwise difficult to capture through traditional unit testing.
The possibilities are numerous, but each of them needs to manage the reproduction of agents accordingly.

The agents can be executed in a playground where they behave like the original end-users.
Reproducing exactly what happened during the tracking phase is infeasible since the concurrency implicitly causes a non-deterministic scenario.
However, the approach is meant to reproduce realistic behaviors, therefore, originating the expected user interactions.
The play engine has to manage the execution of the agents, who observe the reality and detect a status, then decide what actions to perform.
This could be done in real-time if relevant to the testing purposes.
Otherwise, the time in the reality of the playground can be accelerated.
Furthermore, as during the tracking phase agents have learned to statistically infer expectations about changes in the application's state following some action, the target software may also be monitored with regard to its consistency in complying with the expected/past behavior.
At the same time, the opportunity to discover runtime errors remains available.

Optionally, real end-users might take part in the interaction.
Ideally, this should still happen in a dedicated playground, not a production environment.
Moreover, what happens in the playground can be further tracked to generate new and possibly improved data.
In this case, the room for methods and technologies is widely open, and one may choose to leverage reinforcement learning strategies.

\section{Prototype}
\label{sec:case_study}

This section presents the prototype of \emph{Synthetic End-User Testing} developed, and available in the replication package~\cite{replicationpackage}.
This work addresses feasibility of the approach and promotes discussions about future work.
Some key aspects addressed include:
\begin{inparaenum}
    \item user tracking logics to be injected into a codebase without the latter being adapted for this purpose;
    \item the recorded tracking data to be well-formed and useful for the synthesizing of user behavior;
    \item an agent to be enabled to simulate user interactions without adding explicit support for codebase without the latter being adapted for this purpose;
    \item a synthesized agent to rely on past user behavior data to decide what user action to take;
    \item an agent to be constructed to assert the results of an action based on past user behavior data.
\end{inparaenum}

\subsection{System Design}

The subject of this preliminary feasibility study is a small-scale replica of \textsc{Twitter}.
It provides the users with a \emph{GUI} client with which they can connect to a remote server and access a set of functionalities resembling those offered by the popular social platform.
In particular, users can: sign-up, login, logout, view the list of other users registered, choose to follow or unfollow a user, publish new tweets, retweet to and like or unlike any of the tweets displayed, share images (generated) with their tweets and retweets, view their or users the follow's tweets, hence view direct retweets made to their tweets, view who has liked a tweet, and receive push alerts when their tweets have been liked, or a user has started following them.

The application is written in \textsc{Java}, and employs \textsc{Java Swing} for the definition of the \ac{gui} and respects the default \ac{mvc} -based \emph{Observer} design pattern.
Furthermore, the application includes several typical \ac{gui} features such as: asynchronous operations (on most actions, \eg, tweet/retweet, like, menus), pop-up windows, both custom and default \ac{ui} components, as well as graphically reacting on events not promoted by the user events, \eg, receiving alerts.

\subsection{Tracking Layer Injection}

In this prototype, a tracking layer was woven into the client application through \acf{aop}, and required no modification of the codebase.
Instead, a separate \textsc{AspectJ} project was developed.
The tracking layer is in charge of intercepting and logging user actions, injecting hooks for the programmatic triggering of \ac{ui} events, and maintaining an internal representation of the \ac{gui}'s current state and available actions.

It does so by first intercepting declarations of \ac{ui} components to which user action events are added.
Once one of such components is created, the layer appends extra logic that allows it to be notified whenever this becomes the subject of a user action.
Upon receiving such a notification, the layer logs the event and its respective date and application state.
A weak pointer to the components is kept and referenced by an \emph{id}.
Such \emph{id} represents the layout location of the component in the \ac{ui}, as the ordered nesting sequence of its parents, and the text the component carried (when available), thereby making it formal and session-independent.
A numbering system handles the instances in which components list multiple children.
As a result of this naming convention, active components, \ie, currently on screen, are those that are bounded to an \emph{id} starting with the application's window.
Whenever structural \ac{ui} changes occur, these are intercepted, and the \emph{id}s of the action-components are recomputed.
Therefore, the tracker maintains an up-to-date representation of the component carrying actions the user may currently choose to take.
In turn, this allows for the programmatic triggering of user events.
In fact, given the \emph{id} of a target component and the description of a user event (\eg, click), an \textsc{AWT} event is created and invoked on the component.

Finally, for this first prototype, the application state was represented as the name of the view, or page, the user is currently viewing upon taking action.
Declarations and updates of such views are handled similarly to the action components' \emph{id}s.

\subsection{Agents Syntesis}

The concept of \emph{agent} was abstracted to a finite state automata, and later three baseline agent types were derived: \emph{ReplayAgent}, \emph{RandomAgent}, \emph{FrequencyAgent}.
In this particular instance, an agent would:
\begin{inparaenum}
    \item (\emph{Select}) select a user action given the current application state, the set of available actions, and some agent-specific internal logic;
    \item (\emph{Perform}) programmatically trigger the action selected;
    \item (\emph{Await}) obtain the application's state following the performing of the action;
    \item (\emph{Assert}) verify the compliance of the state based on some internal expectation;
    \item return to \num{1}.
\end{inparaenum}

\paragraph{ReplayAgent}
A \emph{ReplayAgent} is the simplest form of agent.
It re-executes the same sequence of actions that some real-world user previously performed, supplied to it in the form of a log file.
Therefore, action selection consists in choosing the next action in the sequence.
Similarly, the state's assertion verifies that the application reached the same condition it did when the user initially performed the same action.
While this design will ensure the agent reproduces realistic action sequences, these are fixed and fails to synthesize user behavior, instead of providing similar value propositions of unit/scenario tests.
It not only leads to practical limitations (\eg, unavailable actions, cyclic operations) but also fails to explore action subsequences that might occur in response to external changes to the application state (\eg, interactions with other agents).

\paragraph{RandomAgent}
A \emph{RandomAgent} is another baseline agent design.
It consists of an agent that selects one of the available actions randomly.
Although it is eventually going to explore more use patterns than a \emph{ReplayAgent}, in its simplest form, it is unable to construct an expectation for the application state following a random action.
Hence, this design also fails to synthesize user behavior and can only unveil implementation level bugs, \eg, unhandled exceptions and error codes.
This strategy seems to diverge from the goals of this project and instead results in an exploration testing strategy with a complete lack of optimization strategy.

\paragraph{FrequencyAgent}
Finally, a \emph{FrequencyAgent} is the only agent introduced in this prototype that makes a baseline attempt to synthesize user behavior.
As the name suggests, this agent assumes the probability of choosing a specific action is only dependent on the application state.
Thus, when supplied with a user's activity log, it builds an action frequency table about each state.
Likewise, it builds a state frequency table about each unique tuple of application state and action taken.
While the first table enables the agent to perform a weighted sampling of an action during the \emph{Select} phase, the second allows the agent to have a weighted expectation about the resulting application state.

\subsection{Play Simulation}

Despite the prototype relying on strong and baseline assumptions, it proved that it is possible to inject into a target software some custom logic for the tracking of user activity and later use this information to synthesize user behavior.
The feasibility of using such agents to simulate end-user activity and therefore test the application was also addressed.
In particular, \emph{FrequencyAgent}s were trained on user activity performed on the target system discussed above, and later two types of bugs were artificially added to the codebase.
The first induced the client application into failing to take the users back to the their feeds (or home) whenever they clicked on an alert of another user liking one of their tweets.
This was designed to occur only after receiving \num{10} alerts, and aimed at asserting whether the agents were able to detect a break in user expectations occurring after some interactions among them.
A second artificial bug introduced in the codebase consisted of a runtime exception being thrown programmatically with a certain probability every time the user requested to follow another user.
Both bugs were eventually identified by the \emph{FrequencyAgent}s.
In the case of the first bug, it correctly reported the weighted expectation of the action following the click on the like alert to consist of the feeds page, but instead of this remaining at the alerts page.
Hence, ultimately confirming that a baseline form of \emph{Synthetic End-User Testing} was compilable for the system specification introduced in this work.

\section{Takeaways}
\label{sec:conclusions}

As a result of the investigations carried out in the compilation of this prototype, a number of observations were made which can inform future development in this context.

The first of these is that the data collected by the tracking layer directly affects how much the agents can learn to synthesize end-user behavior and how effective the agent can be in detecting software behavior breaking user expectations.
The prototype was limited to represent the system state as the name (\emph{id}) of the currently active page.
Despite this being sufficient for detecting errors concerning the updates of pages, it makes it intractable to see any other form of errors, such as, for example, the liking of a tweet failing to update the likes counter.
Therefore, future work should also record richer state representations that enable agents to make finer grain validations.

A second observation concerns the modeling of user actions.
After preliminary qualitative analysis of the performances of the baseline \emph{FrequencyAgent}, it became evident how human decision-makers do not infer action selection based on the current state alone.
Instead, there appears to be an element of sub-flows in using a system, each consisting of multiple actions.
These concern different tasks a user may carry out during an entire session, such as tweeting or retweeting.
\emph{FrequencyAgent} failed to effectively replicate this behavior, with actions found in those sub-flows performed by the real user with equal frequency becoming overlapped.
Future efforts should therefore work towards building end-user models capturing this aspect.
This may be achieved by both computing a more significant state capable of providing an opportunity to infer the actions taken earlier or adding dependency of future actions on those in the past.

As a result, future work should focus on extracting fine-grained information from the tracking phase and leverage deep learning models, such as those above described, to generate agents that can reduce the testing search space by more accurately encoding end-user processes.
Efforts should also evaluate the performance trade-offs of maximizing code coverage against user behavior.

\balance
\bibliography{references, urls}

\end{document}